\documentclass[pra,aps,twocolumn,epsfig,showpacs,superscriptaddress]{revtex4}
\usepackage{amsmath}
\usepackage{amssymb}
\usepackage[dvips]{graphicx}
\usepackage{dcolumn}
\usepackage{bm}
\usepackage[colorlinks=false,dvipdfm]{hyperref} 
\usepackage{setspace}
\usepackage{amsfonts}
\usepackage{rotating}
\usepackage{makeidx}
\usepackage{amsthm}

\usepackage{multirow}
\usepackage{ctable}
\usepackage{booktabs}
\usepackage{threeparttable}
\usepackage{makecell}

\newcommand{\tr}{\mbox{Tr}}

\begin{document}

\title{Most robust and fragile two-qubit entangled states under depolarizing channels \footnote{to appear in Quantum Information $\&$ Computation (QIC)}}

\author{Chao-Qian Pang }
\affiliation{Physics Department, School of Science, Tianjin
University, Tianjin 300072, China}

\author{Fu-Lin Zhang}
\email[Email: ]{flzhang@tju.edu.cn}
\affiliation{Physics Department, School of Science, Tianjin
University, Tianjin 300072, China}

\author{Yue Jiang  }
\affiliation{School of Science, Tianjin Institute of Urban Construction, Tianjin 300384, China}

\author{Mai-Lin Liang}
\affiliation{Physics Department, School of Science, Tianjin
University, Tianjin 300072, China}

\author{Jing-Ling Chen}
\affiliation{Theoretical
Physics Division, Chern Institute of Mathematics, Nankai University,
Tianjin, 300071, China} \affiliation{Centre for Quantum
Technologies, National University of Singapore, 3 Science Drive 2,
Singapore 117543}

\date{\today}

\begin{abstract}
For a two-qubit system under local depolarizing channels, the most
robust and most fragile states are derived for a given concurrence
or negativity. For the one-sided channel, the pure states are proved
to be the most robust ones, with the aid of the evolution equation
for entanglement given by Konrad \emph{et al.} [Nat. Phys. 4, 99
(2008)].
Based on a generalization of the evolution equation for
entanglement, we classify the ansatz states in our investigation by
the amount of robustness, and consequently derive the most fragile
states. For the two-sided channel, the pure states are the most robust
for a fixed concurrence. Under the uniform channel,  the most
fragile states have the minimal negativity when the concurrence is
given in the region $[1/2,1]$. For a given negativity, the most
robust states are the ones with the maximal concurrence, and the
most fragile ones are the pure states with minimum of concurrence.
When the entanglement approaches zero, the most fragile states under
general nonuniform channels tend to the ones in the uniform channel.
Influences on robustness by entanglement, degree of mixture, and
asymmetry between the two qubits are discussed through numerical
calculations.
It turns out that the concurrence and negativity are major factors
for the robustness. When they are fixed, the impact of the mixedness
becomes obvious. In the nonuniform channels, the most fragile states
are closely correlated with the asymmetry, while the most robust
ones with the degree of mixture.

\end{abstract}



\maketitle



\section{Introduction}

Quantum coherent superposition makes quantum information conceptually more powerful
than classical information, which is the essential distinction between a quantum
 system and a classical one. Entanglement is a manifestation of the distinction
 in composite systems \cite{EPR} and is
one of the key resources in the field of quantum information
\cite{Book}. However, unavoidable coupling between a real quantum
system and its environment can cause decoherence, leading to the
destruction of entanglement among subsystems simultaneously.

Because of its important role both in fundamental theory and
applications in quantum information, dynamics of entanglement in a
quantum system under decoherence has attracted wide attention in
recent years.
 And many significant and interesting results have been reported. For instance, the concept of entanglement sudden death
(ESD) has been presented by Yu and Eberly
\cite{yu2004finite,yu2006quantum}, which means that entanglement can
decay to zero abruptly in a finite time while complete decoherence
requires an infinite amount of time. This interesting phenomenon has
been  recently observed in two sophistically designed experiments
with photonic qubits \cite{almeida2007experimental} and atomic band
\cite{laurat2007heralded}. In \cite{konrad2007evolution}, by
utilizing the Jamio{\l}kowski isomorphism, Konrad \emph{et al.}
presented a factorization law for a two qubit system, which
described the evolution of entanglement in a simple and general way,
and has been extended in many directions
\cite{yu2008evolution,li2009evolution,liu2009dynamics,tiersch2008entanglement}.

This paper is concerned with the robustness of entanglement in a
quantum system coupled to noise environments. Vidal and Tarrach
\cite{PhysRevA.59.141} introduced the robustness of entanglement as
a measure of entanglement, corresponding to the minimal amount of
mixing with separable states which washed out all entanglement.
Subsequently, Simon and Kempe \cite{PhysRevA.65.052327} considered
the critical amount of depolarization where the entanglement
vanishes as a quantitative signature of the robustness of the
entanglement, when they studied the robustness of multi-party
entanglement under local decoherence, modeled by partially
depolarizing channels acting independently on each subsystem. This
definition was adopted in the recent work \cite{PhysRevA.82.014301},
in which an interesting residual effect was pointed out on the
robustness of a three-qubit system in an arbitrary superposition of
Greenberger-Horne-Zeilinger state and W state.

We notice that, even in the two-qubit system, there are still some
questions about the robustness to explore. For instance, the results
of \cite{PhysRevA.82.014301} show the robustness increases
synchronously with the degree of entanglement for a two-qubit pure
state. Obviously, it should be very hard to extend this conclusion
to mixed states, because the measures of entanglement for mixed
states are not unique \cite{Wootters97,Wootters98,NEG}, which would
influence the robustness simultaneously . Then, a question arises:
For a given value of entanglement, which states
 are the most robust ones and which are the most fragile ones?
In general, the degree of entanglement in a two-qubit system
 correlates   positively   with the resource of preparation \cite{Wootters97,Wootters98} and its capacity in quantum information \cite{lee2000entanglement,bowen2001teleportation}, therefore it is necessary to compare the stability of states with same entanglement.
Many similar investigations have been reported about ten years.
 For example, the famous \emph{maximally entangled mixed states} \cite{MEMS,MEMS1} can be considered
 to have the most residual entanglement under a global noise channel.
 In \cite{yu2002phonon}, Yu and Eberly studied the robustness and fragility of entanglement in some exactly solvable dephasing
models. In the recent work of Novotn{\`y} \emph{et al.} \cite{novotny2011entanglement}, they present a
 general and analytically solvable decoherence model without any weak-coupling or
 Markovian assumption and distinguish the robust and fragile states.

In the present work, we investigate the most robust entangled states (MRES) and
the most fragile entangled states (MFES) for a given amount of
initial entanglement in the two-qubit system under a local noise channel, and analyze
 connections between the robustness of entanglement and the properties of initial state.
Similar analyses have been done for Werner states in dephasing channel \cite{PhysRevA.76.044101},
 X-states of two two-level atoms in the electromagnetic radiation field \cite{ali2008manipulating}, and dephased pure states in amplitude damping channel \cite{Qian20122931}.
We extend the scope to arbitrary two-qubit entangled states and focus on
 their properties relating to entanglement, and therefore
consider the local depolarizing channels which is invariant under
local unitary operations. In our work, the model in
\cite{PhysRevA.65.052327} and \cite{PhysRevA.82.014301} is
generalized to the nonuniform case. Namely, the coupling strengths
of the two qubits with their environments are different from each
other. This generalization is more  close to practical cases , in
which the entangled two qubits undergo different environments. We
adopted the concurrence \cite{Wootters97,Wootters98} and the
negativity \cite{NEG} as two measures of entanglement for
comparison.
 In Sec. \ref{def}, we review the definitions adopted in this paper
and introduce some notations. In Sec. \ref{onesided},
we selectively give the results of the one-sided channel, which is
an extreme case of the nonuniform channels. To derive the MFES in
this case, a generalization of the \emph{evolution equation for
quantum entanglement} in \cite{konrad2007evolution} is presented.
  In Sec. \ref{twosided}, both the uniform and nonuniform channels are studied. Conclusions and discussions are made in the last section.


\section{Notations and definitions}\label{def}

\subsection{Channels and robustness}

Under local decoherence channels with no interaction between the two
qubits, the dynamics of each qubit is governed by a master equation
depending on its\\ own environment \cite{borras2009robustness}. The
evolution of each reduced density matrix is described by a
completely positive
 trace-preserving map: $\rho_i (t) =\varepsilon_i(t) \rho_i(0),\  i=1,2$, and for the whole state
 $\rho(t)=\varepsilon_1(t)\otimes \varepsilon_2(t) \rho(0)$. In the Born-Markov
approximation these maps (or channels) can be described using its Kraus representation
\begin{eqnarray}\label{Kraus}
\varepsilon_i(t) \rho_i(0)=\sum^{M}_{j=0} E_{ji}\rho_i(0)  E^{\dag}_{ji}
\end{eqnarray}
where $E_{ji}$ satisfying $\sum^{M-1}_{j=0} E_{ji}  E^{\dag}_{ji}=I$, are the Kraus operators, and $M$ is the number of operators
needed to completely characterize the channel. For the depolarizing channels, the Kraus operators are
\begin{eqnarray}\label{DPL}
E_{0i}=\frac{1}{2}\sqrt{3 s_i+1}I_i,\ \   E_{ji}=\frac{1}{2}\sqrt{1- s_i} \sigma_{ji},
\end{eqnarray}
where $\sigma_{ji}\  (j=1,2,3)$ is the $j$th Pauli matrix for
the $i$th qubit. We consider the noise parameter $s_i= \exp(-
\kappa_{i} t)$, where $\kappa_i$ is the decay constant determined by
the strength of the interaction of the $i$th qubit with its
environment and $t$ is the interaction time \cite{dur2004stability}.
 In the Bloch sphere representation for the qubit density operator, the transformation in Eqs. (\ref{Kraus}) and (\ref{DPL})
 is given by $\rho_i (t)= \frac{1}{2}(I_i + s_i \vec{r}_i \cdot \vec{\sigma}_i)$ with $\vec{r}_i$ being the the initial Bloch vector.

Following the method in \cite{PhysRevA.65.052327} and \cite{PhysRevA.82.014301}, we adopt the critical noise parameter
, which is positively associated with the ESD time, as the definition of the robustness.
 The asymmetry between the environments interacting with the two qubits is described by the time-independent parameter $\Delta=(\kappa_1-\kappa_2)/(\kappa_1+\kappa_2)$. It is easy to find that the MRES and MFES for a given value of initial entanglement are
independent of the value of $\kappa_1+\kappa_2$, when the uniform
parameter $\Delta$ is fixed. Therefore, without loss of generality,
we choose the decay constants $\kappa_1=1+\Delta$ and
$\kappa_2=1-\Delta$, with the uniform parameter $\Delta \in [0,1]$.
Then, the noise parameters are written as $s_1= s^{1+\Delta}$ and
$s_2= s^{1-\Delta}$, where $s=e^{-t}$. The robustness for a state
$\rho$ under a nonuniform channel is written as
\begin{eqnarray}\label{RobDef}
\mathcal{R}(\rho)=1-s_{crit}(\rho),
\end{eqnarray}
where $s_{crit}(\rho)$ denotes the critical value of the noise parameter $s$,
 at which ESD of the two-qubit system occurs with the initial state $\rho$.
Though the ESD time is irrelevant to assessing the asymptotic
behavior of the robustness in multi-qubit systems in the limit of
large number of particles
\cite{PhysRevLett.100.080501,zhang2011speed}, it is still
practicable to compare the robustness of entangled states in a given
two-qubit or three-qubit system. In our recent work
\cite{zhang2011speed}, we present the speed of disentanglement as a
quantitative signature of the robustness of the entanglement and
show it is effective to reflect the asymptotic behavior in
multi-qubit systems. One can find that, MRES and MFES in subsection
\ref{RN2} for the uniform channels and the most robust symmetrical
three-qubit pure states in \cite{PhysRevA.82.014301} coincide with
the results in \cite{zhang2011speed}.


\subsection{Measures of entanglement}

%
The concurrence of a pure two-qubit state $| \psi \rangle= c_1 | 00
\rangle +c_2 | 01 \rangle  +c_3 | 10 \rangle  +c_4 | 11 \rangle  $
is given by $\mathcal{C}(| \psi \rangle)=2 |c_1 c_4 - c_2 c_3|$.
 For a mixed state, the concurrence is
defined as the average concurrence of the pure states of the
decomposition, minimized over all decompositions of $\rho =
\sum_{j}p_{j} | \psi_j \rangle \langle \psi_j |$,
\begin{eqnarray}\label{ConM}
\mathcal{C}(\rho)= \min \sum_{j}p_{j}\mathcal{C}(| \psi_j \rangle).
\end{eqnarray}
It can be expressed explicitly as \cite{Wootters97,Wootters98} $\mathcal{C}(\rho)= \max \{0,\sqrt{\lambda_{1}}-\sqrt{\lambda_{2}}-\sqrt{\lambda_{3}}-\sqrt{\lambda_{4}}
\}$,
in which $\lambda_{1},...,\lambda_{4}$ are the eigenvalues of the
operator $R=\rho (\sigma_{y} \otimes \sigma_{y} ) \rho^{*}
(\sigma_{y} \otimes \sigma_{y} )$ in decreasing order and
$\sigma_{y}$ is the second Pauli matrix.

For a bipartite system described by the density matrix  $\rho$,
the negativity is defined as  \cite{NEG,NEG1}
\begin{eqnarray}\label{Neg}
\mathcal{N}(\rho)= 2   \sum_{j} |\lambda_j|,
\end{eqnarray}
where $\lambda_j$ are the negative eigenvalues of $\rho^{T_2}$ and
$T_2$ denotes the partial transpose with respect to the second
subsystem. For the two-qubit system, the partially transposed density
matrix $\rho^{T_2}$ has at most one negative eigenvalue
\cite{MEMS1}.

\subsection{Ansatz states}

To derive MRES and MFES for a given value of concurrence or
negativity, we adopt the approach in \cite{MEMS,MEMS1}. We randomly
generate two-qubit states and derive their degree of entanglement
 and robustness under a series of nonuniform parameter $\Delta \in
\{0,0.1,0.2,...1 \}$.
Plotting them in the corresponding robustness-entanglement planes (such as the ones in Fig. \ref{fig1}),
we fortunately find that, their regions are always the same as the states
\begin{eqnarray}\label{ansatz}
\rho_{ansatz}(r,\theta)= r | \psi (\theta) \rangle \langle \psi (\theta)| + (1-r) |01 \rangle \langle01|,
\end{eqnarray}
where $r \in [0,1]$ and $| \psi (\theta) \rangle = \cos \theta | 00 \rangle + \sin \theta |11 \rangle$ with $\theta \in [0,\pi/2]$.
For each value of $\Delta$, and for both the entanglement measures,
this is supported by $150000$ random states $\rho_{random}$ and $150000$ weighted random states $\rho_{\delta,random}$.
Here, the random states $\rho_{random}$ are uniform in the Hilbert space preserving the Haar measure \cite{NEG,randomU}, and
$\rho_{\delta,random}= (1-\delta)\rho_{ansatz}(r,\theta) + \delta \rho_{random}$ with random parameters $r$, $\theta$ and $\delta$ with the uniform distributions on the interval $[0,1]$, $[0,\pi/2]$ and $[0,0.05]$ respectively.
The numerical results indicate that, the family of states (\ref{ansatz}) contains the entangled states with the maximal or minimal robustness
 for a given degree of entanglement. Therefore, MRES or MFES can be represented in the form of (\ref{ansatz}) with a constraint on its parameters $r$ and $\theta$, which we will derive in the following sections.
We call the state (\ref{ansatz}) the \emph{ansatz state }in this paper.

\begin{figure}
\centering
\includegraphics[width=6.5cm]{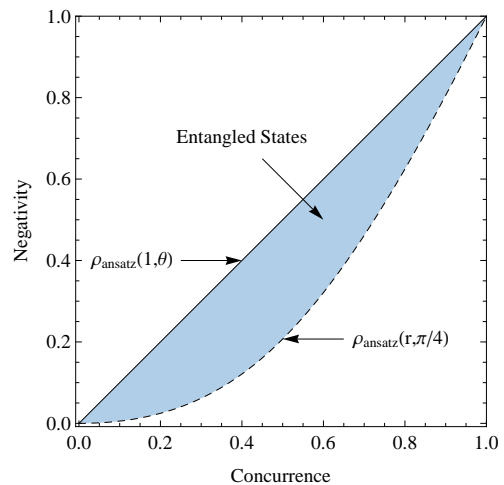} \\
 \caption{(color online) Region of the two-qubit entangled states in the concurrence-negativity plane with the curves of $\rho_{ansatz}(1,\theta)$ (pure state) and $\rho_{ansatz}(r,\pi/4)$. The figure is a slightly
modified version of a figure from \cite{verstraete2001comparison}.}
\label{fig0}
\end{figure}

One point to be mentioned is that, the two extreme cases of the
states correspond to the bounds in the comparison of negativity and
concurrence \cite{verstraete2001comparison}. As shown in Fig.
\ref{fig0}, for a fixed amount of concurrence, the pure state
$\rho_{ansatz}(1,\theta)=| \psi (\theta) \rangle \langle \psi
(\theta)|$ achieves the maximum of negativity, and
$\rho_{ansatz}(r,\pi/4)$ has the minimal negativity. Similarly, when
the negativity is given, the state $\rho_{ansatz}(r,\pi/4)$ is at
the upper bound of concurrence and $\rho_{ansatz}(1,\theta)$ at the
lower one.


\section{One-sided noisy channel}\label{onesided}

\subsection{Robustness vs concurrence in one-sided channel}

 When $\Delta=1$, $s_1=s^2$ and $s_2=1$, the
channel reduces the one-sided depolarizing channel, where only the
first qubit is influenced by its environment.
 The central result of \cite{konrad2007evolution} is the factorization law for an arbitrary pure state $| \psi \rangle$ under a one-sided noisy channel
\begin{eqnarray}\label{factor}
\mathcal{C} [(\$ \otimes I) | \psi  \rangle \langle \psi |] = \mathcal{C} [(\$ \otimes I) | \psi^{+} \rangle \langle \psi^{+}|] \mathcal{C}( | \psi \rangle),
\end{eqnarray}
where $| \psi^{+} \rangle= |\psi(\pi/4)\rangle$ and $ \$ $ denotes
an arbitrary channel not restricted to a completely positive
trace-preserving map. An evident corollary is that,  ESD of systems
set initially in any entangled pure states  occurs at the same time,
and consequently all pure states have the same robustness.

An application of (\ref{factor}) in  \cite{konrad2007evolution}  is the
inequality for mixed states $\rho_0$,
\begin{eqnarray}\label{facRho}
\mathcal{C} [(\$ \otimes I)\rho_0] \leq \mathcal{C} [(\$ \otimes I) | \psi^{+} \rangle \langle \psi^{+}|] \mathcal{C}( \rho_0).
\end{eqnarray}
From (\ref{factor}) and (\ref{facRho}), one can find that under a
given one-sided channel, ESD of a mixed state comes not later than
the pure state with the same concurrence. Hence, MRES with a fixed
concurrence is the pure state. Under the one-sided channel, the
state $|\psi^{+} \rangle$ becomes
 \begin{eqnarray}
 \rho_s =&& \frac{1+s^2}{4}\bigr ( | 00\rangle \langle 00| + |11\rangle \langle11 | \bigr ) \nonumber \\
  &&+ \frac{1-s^2}{4} \bigr( | 01 \rangle \langle 01 | + |10\rangle \langle 10 | \bigr) \nonumber \\
  &&+ \frac{s^2}{2}  \bigr( | 00\rangle \langle 11| + |11\rangle \langle00 |\bigr). \nonumber
 \end{eqnarray}
Its concurrence is $\mathcal{C}(\rho_s) = \max \{(3 s^2-1)/4 ,0 \}$. We can derive the value of robustness
\begin{eqnarray}\label{Rpure}
\mathcal{R}_{MRES}=\mathcal{R}(| \psi  \rangle)=1-\frac{1}{\sqrt{3}},
\end{eqnarray}
which is a constant and independent of the concurrence.

\begin{figure}
\centering
\includegraphics[width=8cm]{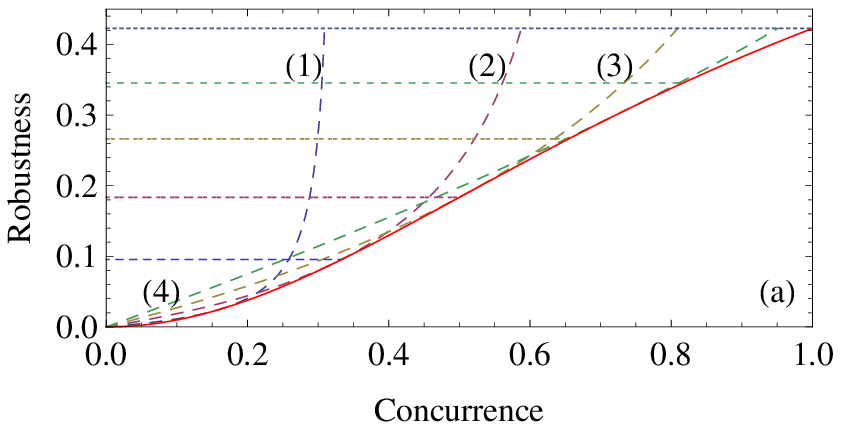} \\
\includegraphics[width=8cm]{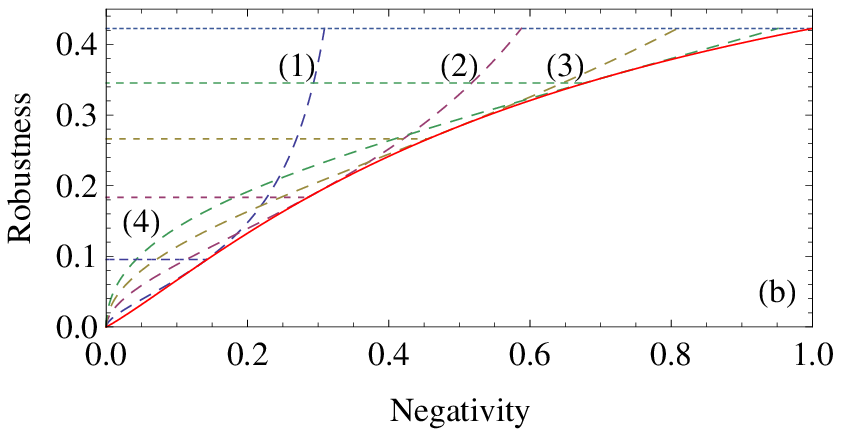}\\
 \caption{(color online) Plots of the states $\rho(c,p)$ in the (a) robustness-concurrence plane and (b) robustness-negativity plane under one-sided noise channel. The dotted lines correspond to the parameter $c=1,0.8,0.6,0.4,0.2$ from top to bottom, in both (a) and (b). The dashed lines refer to the parameter (1) $p=\tan(0.1\times \pi/2)$, (2) $p=\tan(0.2\times \pi/2)$, (3) $p=\tan(0.3\times \pi/2)$ and (4) $p=\tan(0.4\times \pi/2)$.
 The solid lines are the MFES in the two cases.}
\label{fig1}
\end{figure}

To obtain MFES, we generalize the factorization law
(\ref{factor}) to the arbitrary two-qubit states. Namely, if two
states satisfy
\begin{eqnarray}\label{rho1rho2}
\rho_1 = \gamma (I\otimes M) \rho_0 (I\otimes M^{\dag})
\end{eqnarray}
with $1/\gamma = \tr [(I\otimes M) \rho_0 (I\otimes M^{\dag})] $  and $M= I + \vec{a} \cdot \vec{\sigma},\ (a\in[0,1])$,
the relation
\begin{eqnarray}\label{facArb}
\mathcal{C} [(\$ \otimes I) \rho_1 ] \mathcal{C} ( \rho_0 )  = \mathcal{C} [(\$ \otimes I) \rho_0 ] \mathcal{C} ( \rho_1 )
\end{eqnarray}
holds for all one-sided channels $\$ $.
We omit the proof of the above relation, which is exactly the same
as the process for (\ref{factor}) in \cite{konrad2007evolution}. As
for the pure states, we can conclude that the robustness of $\rho_0$
and $\rho_1$ in (\ref{rho1rho2}) have the same value.

Then, we find that the ansatz states (\ref{ansatz}) can be
classified by the amount of robustness, with the aid of the
generalized factorization law (\ref{facArb}). Namely, choosing $
\rho_0=  \rho_{ansatz}(c,\pi/4)$ and $\vec{a}=(0,0,(1-p)/(1+p))$,
the ansatz states are written as
\begin{eqnarray}\label{rho1rho2cp}
 \rho_{ansatz}(r, \theta)=\rho (c,p) = \gamma (I\otimes M) \rho_0 (I\otimes M^{\dag}),
\end{eqnarray}
where the relations between the two pairs of parameters are
\begin{eqnarray}\label{cp}
c=\frac{r-r\cos 2\theta}{1-r \cos 2\theta },\ \ \ \ \ \ p=\tan \theta.
\end{eqnarray}
The robustness of the ansatz state depends only on the parameter $c$ in $\rho_0$ and is given by
\begin{eqnarray}\label{Rcp}
\mathcal{R} [\rho (c,p)]=1-\sqrt{\frac{2-c}{2+c}},
\end{eqnarray}
which which becomes the result for the pure states in
(\ref{Rpure}) when $c=1$. The concurrence is obtained as
\begin{eqnarray}\label{Ccp}
\mathcal{C} [\rho (c,p)]=\frac{2cp}{c+(2-c)p^2} =  r \sin 2 \theta.
\end{eqnarray}
In Fig. \ref{fig1} (a), we plot the ansatz states with some discrete values of $c$ or $p$ in the robustness-concurrence plane.
One can find that MFES achieve the maximal concurrence when the value of robustness is fixed. It is easy to determine the maximal value of concurrence in (\ref{Ccp}) with a fixed $c$.
Then the constraint on the ansatz sates for the MFES can be obtained as
\begin{eqnarray}\label{MFC1cp}
c(1+p^2)-2p^2=0,
\end{eqnarray}
where $0\leq p \leq 1$, or equivalently
\begin{eqnarray}\label{MFC1rt}
2r\cos^2  \theta=1,
\end{eqnarray}
with $0\leq \theta \leq \pi/4$ and $1/2 \leq r \leq 1$. It is
interesting that the conditions leading to the length of the Bloch
vector for the two qubit are $r_1=2(1-r)$ and $r_2=0$, and their
difference $\delta_{r}=r_1-r_2$ satisfies $\delta_r +
\mathcal{N}=1$. Our numerical result shows that, MFES maximize the
difference $\delta_{r}$ with a fixed value of concurrence.


\subsection{Robustness vs negativity  in one-sided channel}

It is easy to prove that MRES under an arbitrary one-sided noise
channel for a given negativity
 are also pure states, with the aid of the results in above subsection.
This conclusion can be obtained by supposing $\rho$ denotes an arbitrary entangled two-qubit state, $| \psi_1  \rangle$ and $| \psi_2  \rangle$ are two pure states, and
$\mathcal{N}(\rho)=\mathcal{N}(| \psi_1  \rangle)$ and $\mathcal{C}(\rho)=\mathcal{C}(| \psi_2  \rangle)$.
 Two relations can be obtained as $\mathcal{R}(\rho)\leq\mathcal{R}(| \psi_2  \rangle)$ and $\mathcal{R}(| \psi_1  \rangle)= \mathcal{R}(| \psi_2  \rangle)$, from which we deduce the conclusion.

In Fig. \ref{fig1}, one can notice that the region of the ansatz
states in the robustness-negativity plane and the one in the
robustness-concurrence case are similar. Thus, in order to determine
MFES we can just derive the states $\rho(c,p)$ with the maximal
negativity for a fixed $c$. The negativity of the  ansatz states is
\begin{eqnarray}\label{Ncp}
\mathcal{N} [\rho_{ansatz}(r, \theta)]=\sqrt{r^2 \sin^2 2\theta+(1-r)^2}-(1-r).
\end{eqnarray}
Inserting the relations (\ref{cp}) into Eq. (\ref{Ncp}), one can find that the maximal negativity occurs when
\begin{eqnarray}\label{CPN}
p = \sqrt{\frac{2 \sqrt{2-c} (-1+c) c^{3/2}+2 c^2-c^3}{-8+24 c-20 c^2+5 c^3}}.
\end{eqnarray}
The corresponding relation of the parameters $r$ and $\theta$ is very close to the numerical results for $\Delta=0.99$ in Fig. \ref{fig5} (a) in the following section.

In the MFES, when the parameter $c \rightarrow 1$, the entanglement
$\mathcal{N}  \rightarrow 1+2(c-1)$
  and the difference between the lengthes of the two Bloch vectors $\delta_r$ approaches $2(1-c)$.
   In other words, MFES with negativity close to $1$ reaches the maximal $\delta_r$.
 At the other extreme, when the parameter $\theta \rightarrow 0$, and meanwhile entanglement of the entangled composition decreases, its proportion $r \rightarrow 1$.
 However, the MFES for concurrence given in (\ref{MFC1cp}) and (\ref{MFC1rt}), the proportion $r$ drops to $1/2$ when $\theta \rightarrow 0$.
Now, MFES have the maximal entropy among the family of the ansatz states.

\section{Two-sided noisy channel}\label{twosided}
\begin{figure}
\centering
\includegraphics[width=8cm]{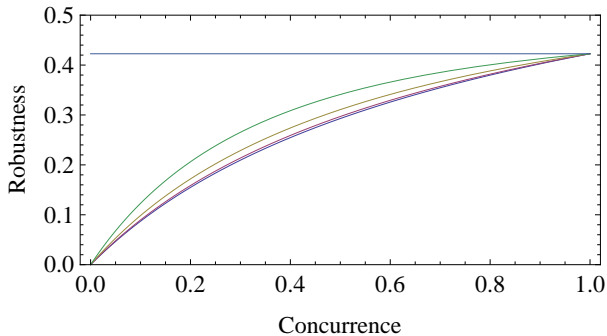} \\
 \caption{(color online) The relation between the robustness and the concurrence
of a two-qubit system in a pure entangled state. From top to bottom the
curves refer to $\Delta=1,0.75,0.5,0.25,0$.}
\label{fig2}
\end{figure}


We fail in generalizing the relations in (\ref{rho1rho2}) and (\ref{facArb}) to the two-sided channel case.
Therefore, we calculate the ESD critical condition of the ansatz states directly
\begin{eqnarray}\label{ESD}
&&\mathcal{P}=4r^2 \sin^2 2\theta s_1^2 s_2^2  -[1+(1-2r)s_1 s_2]^2  \nonumber\\
&& \ \ \ +[r \cos 2\theta (s_1-s_2)+(1-r)(s_1+ s_2)]^2=0,
\end{eqnarray}
which is one of the main bases to determine MRES and MFES in
this section. When $r=1$, one can obtain the relation between the
entanglement of pure states and their critical noise parameters
\begin{eqnarray}
\mathcal{C}^2 (| \psi \rangle)=\mathcal{N}^2 (| \psi \rangle) = \frac{(1-s_1^2)(1-s_2^2)}{4s_1^2 s_2^2 -(s_1-s_2)^2}.
\end{eqnarray}
For the Bell state $|\psi^+ \rangle $, the above equation gives the
robustness $\mathcal{R}(|\psi^+ \rangle)= 1-1/\sqrt{3}$, which is
independent of the nonuniform parameter $\Delta$. In this sense, our
generalization of robustness is reasonable and rational. When
$\Delta=0$, $\mathcal{R}(|\psi \rangle)= 1-1/\sqrt{2\mathcal{C} (|
\psi \rangle)+1}$, which is the result of the uniform channels given
in \cite{PhysRevA.82.014301}. In Fig. \ref{fig2}, we plot the
robustness and the entanglement of the pure state. Under a fixed
$\Delta$,
 the robustness is a monotone increasing function of the entanglement.
 A non-maximally entangled pure state becomes more fragile as $\Delta$ decreases.


\subsection{Robustness vs concurrence in two-sided channel}
Because of the invariance of the depolarizing channels under LU transformations,
 we can show that MRES for concurrence under the two-sided channel are the pure states, which has the maximal negativity for a fixed concurrence as shown in Fig. \ref{fig0}. According to the procedure given by Wootters
\cite{Wootters98}, one can always obtain a decomposition $\{ |
\phi_i \rangle \}$
 minimizing the average concurrence in Eq. (\ref{ConM}), $\rho_0= \sum_i t_i
| \phi_i \rangle \langle \phi_i |$, in which $\sum_i t_i =1$ and all
the elements $| \phi_i \rangle \langle \phi_i |$ have the same value of concurrence as the mixed state
$\rho_0$.
The elements are equivalent under LU transformation to the same
state $| \psi\rangle$ 
\begin{eqnarray}
| \phi_i \rangle \ =U^A_i \otimes U^B_i | \psi \rangle,
\end{eqnarray}
with the concurrence $\mathcal{C}(| \psi \rangle)=\mathcal{C}(
\rho_0)$. After passage through a basis-independent local channel $
\$_1 \otimes \$_2 $, the concurrence  $\mathcal{C}[ (\$_1 \otimes
\$_2)| \phi_i \rangle \langle \phi_i|  ]= \mathcal{C}[ (\$_1 \otimes
\$_2)| \psi \rangle \langle  \psi |]$. It then immediately follows,
by convexity, that
\begin{eqnarray}
\mathcal{C}[ (\$_1 \otimes \$_2)  \rho_0  ] \leq \mathcal{C}[ (\$_1 \otimes \$_2)| \psi
\rangle \langle  \psi |],
\end{eqnarray}
which proves that MRES for fixed concurrence are the pure states.

\begin{figure}
\centering
\includegraphics[width=8.7cm]{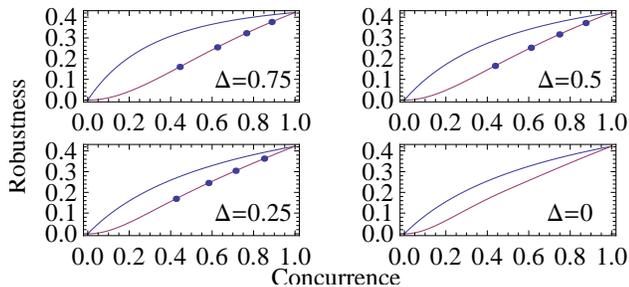} \\
 \caption{(color online)  The relation between the robustness and the concurrence of the MRES and MFES under different nonuniform parameters, $\Delta=0.75,0.5,0.25,0$. The dots on the lines denote the quasi-MFES, whose parameters $\beta=0.1,0.2,0.3,0.4$, with the values of concurrence increasing .}
\label{fig3}
\end{figure}

Similar to the case of the one-sided depolarizing channels,
 MFES in the present case are also the states achieving the maximal concurrence for a given value of robustness, which can be noticed in Fig. \ref{fig3}.
 From (\ref{Ccp}) and (\ref{ESD}), we find that for a given pair of the critical noise paraments $s_1$ and $s_2$,
 or equivalently the robustness $\mathcal{R}$ and the nonuniform parameter $\Delta$, MFES can be obtained by applying the restraint
\begin{eqnarray}\label{alphabeta}
\alpha =\frac{1}{4} \biggr[ 1+ \sqrt{8 \Omega (2 \beta -1) \beta +1} \biggr]
\end{eqnarray}
on the ansatz states (\ref{ansatz}), where $\alpha = r \cos^2 \theta \in [0,1/2]$, $\beta = r \sin^2 \theta \in [0,1/2]$ and $\Omega=\frac{s_1^2 (1-s_2^2)}{s_2^2 (1-s_1^2)}$.
 When $\Delta=1$, the parameter $\Omega=0$ and the restraint reduces to $\alpha =
 1/2$, which is precisely the result of the one-sided channels in (\ref{MFC1cp}) and (\ref{MFC1rt}).
%

For the uniform noise channels where $\Delta=0$ and $\Omega=1$, the relation in (\ref{alphabeta}) becomes $\alpha=1/2 - \beta$ when $\beta\leq 1/4$, and $\alpha=\beta$ when $\beta>1/4$.
Consequently, the MFES can be written in two corresponding  regions as
\begin{eqnarray}
\rho_{MFES}= \left \{
\begin{array}{lr}
\frac{1}{2} [| \psi (\theta) \rangle \langle \psi (\theta)| +  |01 \rangle \langle01|] ,      \; &  \mathcal{C} < 1/2 \;, \\
r  | \psi^+\rangle \langle \psi^+| + (1-r) |01 \rangle \langle01|, \; &  \mathcal{C} \geq 1/2  \;.
\end{array}
\right.
\end{eqnarray}
In the region  $ \mathcal{C} \geq 1/2$, $r\geq 1/2$ and
$\theta=\pi/4$ , the MFES are the states $\rho_{ansatz}(r,\pi/4)$
with the minimal negativity for a fixed concurrence
\cite{verstraete2001comparison} (see Fig. \ref{fig0}).
 In the other, where $\mathcal{C} < 1/2$, $r=1/2$ and $\theta\in[0,\pi/4)$,
 MFES have the maximal mixedness among the ansatz states.
Here, we adopt the linear entropy $\mathcal{S}_L(\sigma)=
\frac{4}{3}(1-\tr \sigma^2)$ as the measure of mixedness, and for the ansatz states
$\mathcal{S}_L(\rho_{ansatz})= \frac{8}{3}(1-r)r$. One can notice
that, MFES in this region satisfy $\mathcal{S}_L(\rho_{MEMS}) \geq
\mathcal{S}_L(\rho_{MFES}) \geq
\mathcal{S}_L[\rho_{ansatz}(r,\pi/4)]$ and $\mathcal{N}(\rho_{MEMS})
\geq \mathcal{N}(\rho_{MFES}) \geq
\mathcal{N}[\rho_{ansatz}(r,\pi/4)]$, where $\rho_{MEMS}$ are the
\emph{maximally entangled mixed states} in \cite{MEMS}.



When $0< \Delta <1$, MFES for the arbitrary depolarizing channels is
given by the solution of equations (\ref{ESD}) and
(\ref{alphabeta}), which is shown in Fig. \ref{fig3} for some values
of $\Delta$. It can be noticed that, the difference between the
robustness of MRES and MFES decreases with the decrease in the
nonuniform parameter $\Delta$. When $\beta\rightarrow0$, the
concurrence $\mathcal{C}\rightarrow0$, the proportion of the
entangled state in MFES approaches $1/2$.

 Considering the absence of a brief expression of MFES, we present a family of quasi-MFES. Namely, we replace the critical depolarizing parameter $s^2 \in [1/3,1]$ in  (\ref{alphabeta}) by its mean value $2/3$, and obtain  $\Omega\rightarrow\widetilde{\Omega}=\frac{(2/3)^{\Delta-1}-1}{(2/3)^{-\Delta-1}-1}$.
  Obviously $\widetilde{\Omega}|_{\Delta=0}=1$ and $\widetilde{\Omega}|_{\Delta=1}=0$. The positions of the quasi-MFES in the robustness-concurrence plane are shown in Fig. (\ref{fig3}), which are very close to the MFES. Our numerical result shows, for a given $\beta$, the fidelity of the quasi-MFES and the MFES $F(\rho_{quasi-MFES},\rho_{MFES}) \geq 1-10^{-4}$ with $F(\rho_1,\rho_2)= \bigr[\tr \bigr( \sqrt{\sqrt{\rho_1} \rho_2
\sqrt{\rho_1}} \bigr) \bigr]^2$, which is supported by one million
pairs of randomly generated quasi-MFES and MFES.







\subsection{Robustness vs negativity  in two-sided channel}\label{RN2}

\begin{figure}
\centering
\includegraphics[width=8cm]{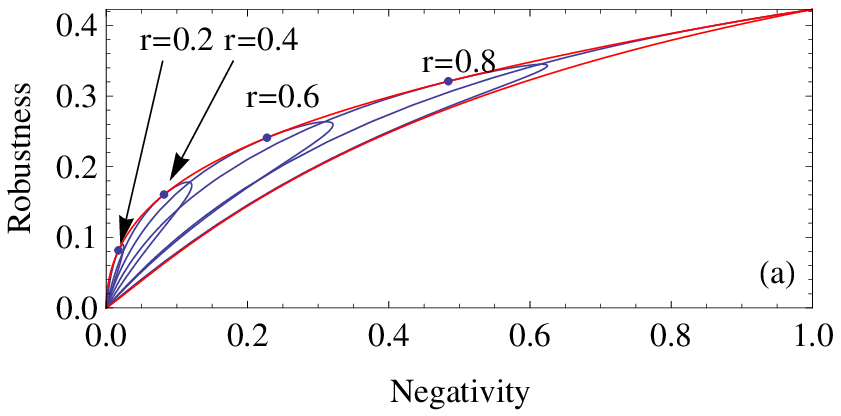} \\
\includegraphics[width=8cm]{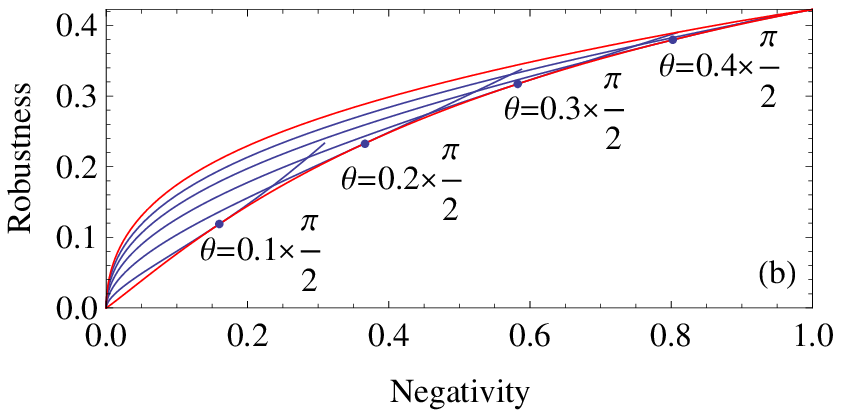} \\
 \caption{(color online) Plots of the states $\rho(r,\theta)$ in the robustness-negativity planes with the MRES and MFES when the nonuniform parameters $\Delta=0.5$. The parameters $r$/$\theta$ takes a set of discrete values in (a)/(b), which are labeled nearby the tangent points with the MRES/MFES.}
\label{fig4}
\end{figure}

The MRES or MFES for the negativity under the two-sided depolarizing channels are also the states which minimize or maximize the negativity for a fixed robustness. We adopt the method of Lagrange multiplier to derive the extremum of the negativity in (\ref{Ncp}) under the constraint (\ref{ESD}), where $s_1$ and $s_2$ are constants. It is obtained that the parameters $r$ and $\theta$ satisfy the the condition,
\begin{eqnarray}\label{NPPN}
\frac{\partial \mathcal{N}}{\partial r} \frac{\partial \mathcal{P}}{\partial \theta} - \frac{\partial \mathcal{N}}{\partial \theta} \frac{\partial \mathcal{P}}{\partial r} =0,
\end{eqnarray}
when the negativity of an ansatz state reaches an extremum for a
fixed robustness. Forms of MRES and MFES are given by the physically
accepted solutions to the Eqs. (\ref{ESD}) and
(\ref{NPPN}).

When $\Delta=0$, one can check that the two solutions $\theta=\pi/4$ and
$r=1$ correspond to MRES and MFES respectively. In other words,
under the uniform channel, MRES are
$\rho_{MRES}=\rho_{ansatz}(r,\pi/4)$, and MFES
$\rho_{MFES}=\rho_{ansatz}(1,\theta)$ are the pure states. As shown
in Fig. \ref{fig0}, for a given value of negativity,
$\rho_{ansatz}(r,\pi/4)$ has the maximum of concurrence and the pure
state $\rho_{ansatz}(1,\theta)$ has the minimum. This result and the
corresponding one for concurrence indicate that, in the uniform
channel, when one of the two degrees of entanglement is fixed, the
other is the main factor affecting the robustness.


\begin{figure}
\centering
\includegraphics[width=8cm]{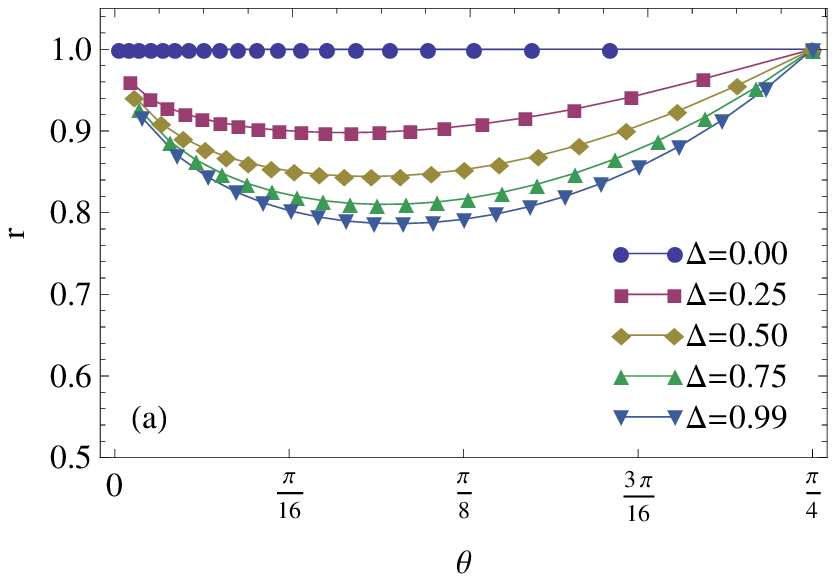} \\
\includegraphics[width=8cm]{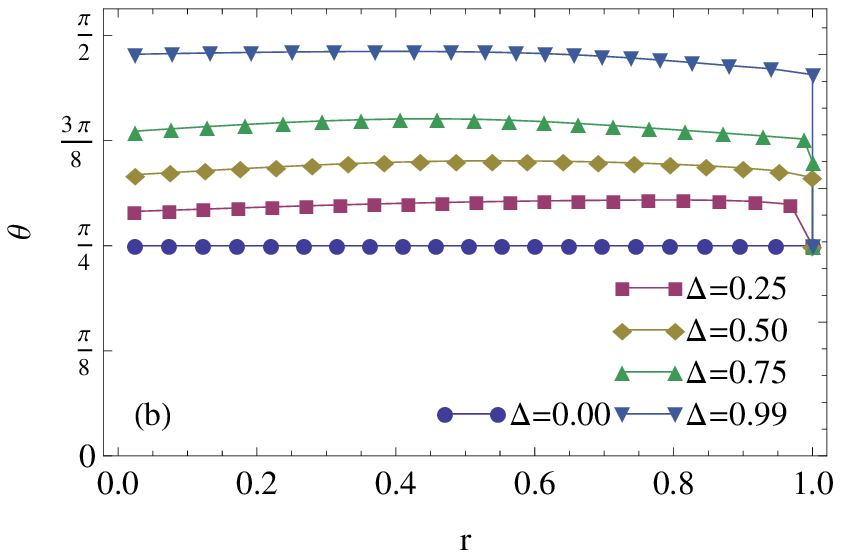}\\
 \caption{(color online) Relations of the parameters $r$ and $\theta$ for the MFES (a) and MRES (b) with different inhomogeneous parameters, $\Delta=0.99,0.75,0.5,0.25,0$. The robustness takes a set of regularly spaced values.}
\label{fig5}
\end{figure}

We were not able to derive an explicit expression of the solution to
(\ref{ESD}) and (\ref{NPPN}) for arbitrary nonuniform parameter $\Delta \in (0,1)$.
Thus we plot the ansatz states in the robustness negativity plane
for various values of the nonuniform parameter $\Delta$, one of
which is shown in Fig. \ref{fig4}. One can notice that, MRES can be
expressed as the ansatz states with a constrain as $\theta=
\theta(r)$, with $r$ varying from $0$ to $1$. And the constrain for
MRES is  $r= r(\theta)$ , where the domain is $\theta \in
[0,\pi/4]$. Then we search the constrains by numerical computation,
and plot them in Fig. \ref{fig5}. It is obvious that the proportion
of the entangled composition $r$ in MFES is always more than $3/4$.
 The proportion approaches $1$ when the entanglement $\mathcal{N} \rightarrow 0$, and
 increases with decreasing $\Delta$.
Under a fixed $\Delta$, the parameter $\theta$ seldom fluctuates when the robustness varies in a very large region.
When $r$ approaches $1$, $\theta$ decreases to $\pi/4$ quickly, and the robustness tends to the maximum $1-1/\sqrt{3}$.
 When $\Delta \rightarrow 1$, the domain of the robustness becomes a point $\mathcal{R}=1-1/\sqrt{3}$, where $r=1$ and $\theta$ varies from $\pi/2$ to $\pi/4$.

\section{Conclusion and Discussion}\label{concl}
In conclusion, we have investigated the robustness of entangled
states for two-qubit system under local depolarizing channels. MRES
and MFES are derived for a given amount of entanglement measured by
concurrence and negativity. With a numerical simulation, we find a
family of ansatz states $\rho_{ansatz}(r,\theta)$ in (\ref{ansatz}),
which contains MRES and MFES.
 In Table \ref{tab1}, we list the constrains on the parameters $r$ and $\theta$ corresponding to MRES and MFES in the one-sided channel and the uniform channel.

\renewcommand\arraystretch{1.5}
\begin{table}[!hbp]
\begin{tabular}{l | c | c |c | c }
\Xhline{1pt}
\multirowthead{2}{ }&
\multicolumn{2}{c |}{$\Delta=0$} & \multicolumn{2}{c}{$\Delta=1$}\\
\cline{2-5}
& MRES & MFES & MRES & MFES    \\
\hline
$\mathcal{C}$ & $r=1$ &\thead{$r=\frac{1}{2}$ for $\mathcal{C}<\frac{1}{2}$  \\ $\theta=\frac{\pi}{4}$ for $\mathcal{C}\geq \frac{1}{2}$ }  & $r=1$ & $2r \cos^2 \theta=1$ \\
\hline
$\mathcal{N}$  & $\theta=\frac{\pi}{4}$ & $r=1$ & $r=1$ & \thead{Defined by \\ Eqs. (\ref{CPN}) and (\ref{cp})}   \\
\Xhline{1pt}
\end{tabular}
\caption{The constrains on the parameters in the ansatz state $\rho_{ansatz}(r,\theta)$ when it achieves the MRES or MFES for a fixed value of concurrence $\mathcal{C}$ or negativity $\mathcal{N}$  in the one-sided channel ($\Delta=1$) and the uniform channel ($\Delta=0$).}\label{tab1}
\end{table}
\renewcommand\arraystretch{1}

Under the one-sided channel, for both measures of entanglement the
pure states are proved to be MRES by utilizing the factorization law
for the evolution of concurrence given by
\cite{konrad2007evolution}. A generalized factorization law is
presented in (\ref{rho1rho2}) and (\ref{facArb}). Based on this
result, we classify the ansatz states by the values of robustness
and derive MFES for the two measures of entanglement. In MFES for
concurrence, the length of the Bloch vector for the free qubit is
zero, and
 the difference between the lengthes of the two Bloch vectors, $\delta_r$, reaches its maximum for a given concurrence.
 When the value of negativity approaches $1$, the corresponding  MFES approach the states with the maximal $\delta_r$ for a fixed negativity.

\begin{figure}
\centering
\includegraphics[width=8cm]{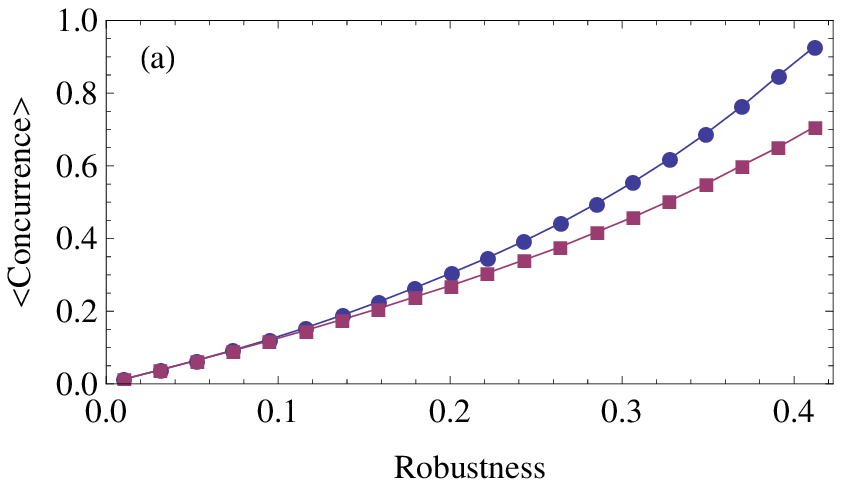} \\
\includegraphics[width=8cm]{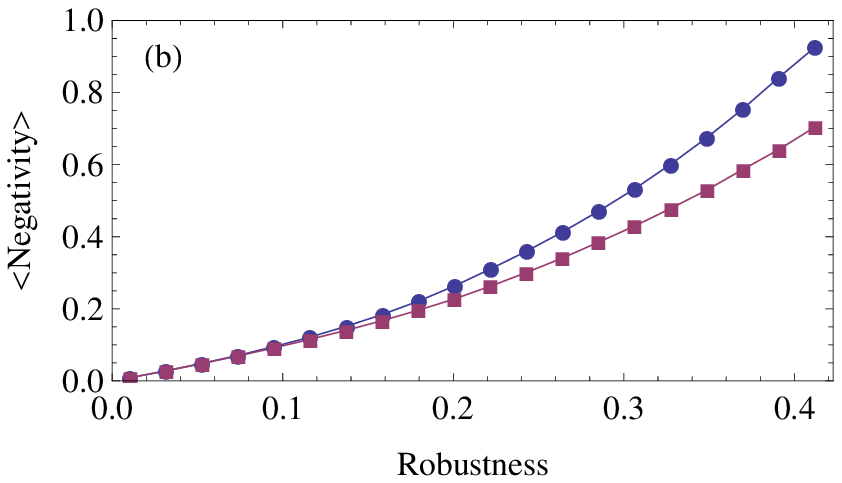}\\
\includegraphics[width=8cm]{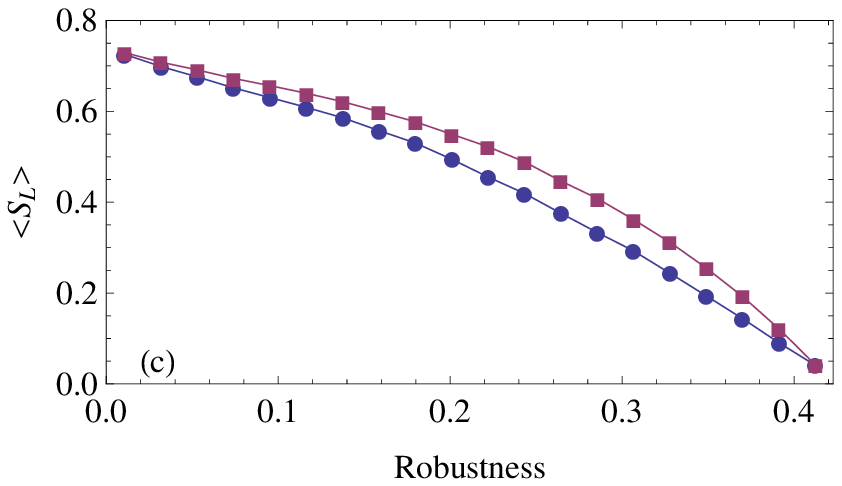}\\
 \caption{(color online) Averages of (a) concurrence, (b) negativity and (c) linear entropy $S_L$ of the states with a given amount of robustness in the case of $\Delta=0$ (dots) and $\Delta=1$ (squares).}
\label{fig6}
\end{figure}

Under the uniform two-sided channel, MFES for a given negativity are
the pure states, and MRES are $\rho_{ansatz}(r,\pi/4)$.
 In contrast, MRES for concurrence are the pure states, while MFES are
$\rho_{ansatz}(r,\pi/4)$, when $\mathcal{C} \geq 1/2$, but the
states $\rho_{ansatz}(1/2,\theta)$ with $\theta\in[0,\pi/4)$ when
$\mathcal{C} < 1/2$. The pure states and states
$\rho_{ansatz}(r,\pi/4)$ are precisely on the boundaries of the
region for arbitrary two-qubit entangled states in the
concurrence-negativity plane \cite{verstraete2001comparison}, which
is shown in Fig. \ref{fig0}. The states $\rho_{ansatz}(1/2,\theta)$
have the maximal linear entropy among the ansatz states, and both
its negativity and linear entropy are between the ones of
$\rho_{ansatz}(r,\pi/4)$ and $\rho_{MEMS}$, when they have the equal
concurrence. Under general two-sided channels, the pure states are
proved to be most robust when the concurrence is given. When the
entanglement approaches zero, MFES for the two entanglement measures
tend to the results in the uniform channel.

\begin{figure}
\centering
\includegraphics[width=8cm]{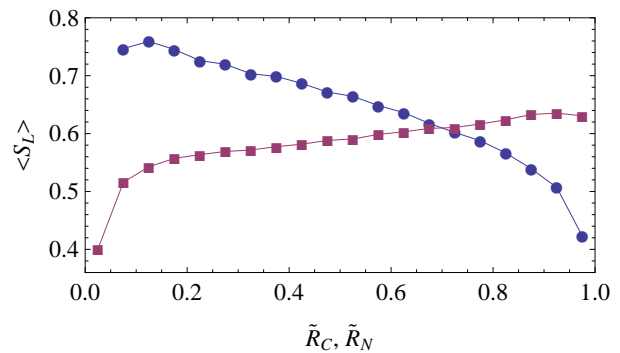} \\
 \caption{(color online) Averages of linear entropy $S_L$ of the states a given amount of normalized robustness $\tilde{\mathcal{R}}_{C}$ (dots) and $\tilde{\mathcal{R}}_{N}$ (squares) when $\Delta=0$.}
\label{fig7}
\end{figure}
\begin{figure}
\centering
\includegraphics[width=8cm]{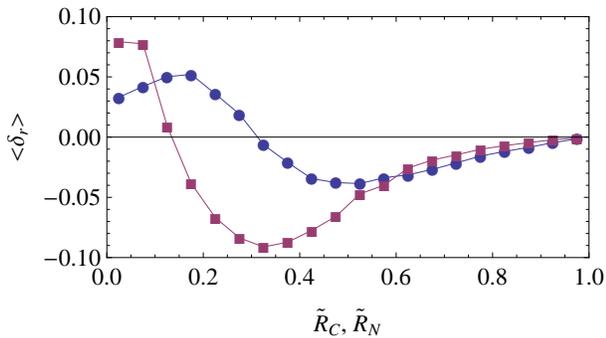} \\
 \caption{(color online) Averages of $\delta_r$ of the states a given amount of normalized robustness $\tilde{\mathcal{R}}_{C}$ (dots) and $\tilde{\mathcal{R}}_{N}$ (squares) when $\Delta=1$.}
\label{fig8}
\end{figure}

To make a interpretation to the characteristics of the MRES and MFES mentioned above, we
analyze the influence on the robustness caused by the entanglement properties and other possible factors based on $300000$ random entangled states.
We adopt the decomposition of the random states in \cite{NEG} as $\rho_{random}=U \rho_D U^{\dag}$, and generate the $4\times4$ unitary matrices $U$ uniformly
under the Haar measure \cite{randomU}.
The difference from \cite{NEG} is that we generate three independent
numbers $\alpha_j, (j=1,2,3)$ uniformly in the interval $[0,\pi/2]$ and get the nonzero elements of the diagonal density matrices
 $\rho_D$ as $\{\cos^2 \alpha_1 \cos^2 \alpha_2$,$\cos^2 \alpha_1 \sin^2 \alpha_2$, $\sin^2 \alpha_1 \cos^2 \alpha_3$, $\sin^2 \alpha_1 \sin^2 \alpha_3 \}$.
In this scenario, without affecting the qualitative conclusions in the following paragraphs,  the probability of the states nearby the MFES and MRES is increased.
%
%
Based on the random entangled states, the averages of some quantities of the states with the given values of the robustness or normalized robustness are plotted in Figs. \ref{fig6},\ref{fig7}, and \ref{fig8}.
 Here, the normalized robustness $\tilde{\mathcal{R}}_{C}$ and $\tilde{\mathcal{R}}_{N}$ are defined as
\begin{eqnarray}
\tilde{\mathcal{R}}_{C,N}(\rho)=\frac{\mathcal{R}(\rho)-\mathcal{R}(\rho_{MFES})}{\mathcal{R}(\rho_{MRES})-\mathcal{R}(\rho_{MFES})},
\end{eqnarray}
where $\rho_{MRES}$ and $\rho_{MFES}$ are the MRES and MFES with the same concurrence or negativity as $\rho$, corresponding to the subscripts $C$ and $N$ respectively.


In Fig. \ref{fig6}, one can find that, in the statistical sense, the
robustness increases with the entanglement, measured by concurrence
and negativity, but decreases with the degree of mixture.
The trend of the line for $\tilde{\mathcal{R}}_{C}$ in Fig.
\ref{fig7} is in accord with Fig. \ref{fig6}(c), but the one for
$\tilde{\mathcal{R}}_{N}$ shows an opposite tendency. These indicate
two points: (1) When the value of negativity is given, the
concurrence is the dominating factor for the robustness in the
uniform channel, and the trend of the linear entropy results from
the changes of concurrence; (2) For the case with a fixed
concurrence,
 the degree of mixture may be a influential factor of the robustness besides the negativity.
These agree with the MRES and MFES in the uniform channel. For a
fixed negativity, MRES maximize the concurrence and MFES minimize
it. For a fixed concurrence, MRES maximize the negativity and
minimize the linear entropy, and MFES minimize the negativity in the
region of $\mathcal{C} \geq 1/2$ but has the maximal linear entropy
among the ansatz states for $\mathcal{C} < 1/2$. And both the
negativity and the linear entropy of  MFES for $\mathcal{C} < 1/2$
are between the MEMS and the states minimizing the negativity.


One can compare the results for $\Delta=0$ and $\Delta=1$ in Fig.
\ref{fig6} and find that when the noise channels become nonuniform,
the influence of the entanglement on the robustness decrease, and
the relationship between the linear entropy and the robustness
weakens for $\mathcal{R}$ close to zero but strengthens near its
maximum. It is obvious that the robustness in the nonuniform channel
is affected by the unbalance between the two qubits. Here, we
consider $\delta_r$ as a quantitative signature of the asymmetry. It
is shown in Fig. \ref{fig8} that, in the one-sided noise channel,
when the value of concurrence or negativity is fixed, the
relationship between $\delta_r$ and robustness is significant when
the normalized robustness is less than $0.4$, but becomes less
evident when the robustness approaches its maximum.
Therefore, we surmise that, as the channel becomes nonuniform, MFES
are closely correlated with $\delta_r$, but MRES with the linear
entropy. This point is well represented in the MRES and MFES. When
the amount of concurrence is given, MRES under general two-sided
channels maximize the negativity and minimize the linear entropy,
 and MFES in the one-sided channel achieve the maximum of $\delta_r$.
When the negativity is fixed, the MRES changes from the states with
maximal concurrence to the states minimizing the linear entropy as
the nonuniform parameter $\Delta$ increases from $0$ to $1$. The
MFES for negativity in the one-sided channel have the maximal
$\delta_r$ when $\mathcal{N} \rightarrow 1$.

At last, we make some prospects for further extension of the results
in this paper. Based on the classification of the ansatz states
under one-sided channel, it is easy to study the relation between
robustness and entanglement. It leaves us a question of whether the
arbitrary two-qubit states can be classified by robustness? The
answer to this question would bring us an explicit expression of
robustness under one-sided channel. The evolution equation of
entanglement given in \cite{konrad2007evolution} has been
generalized in many directions
\cite{yu2008evolution,li2009evolution,liu2009dynamics,tiersch2008entanglement}.
 As an application of the generalized
evolution equations, it is interesting to study robustness of
entanglement in multi-qubit or multi-qudit systems.
  On the other hand,
 the capacity of an entangled qubit pair in quantum information \cite{lee2000entanglement,bowen2001teleportation} does not depend on the amount of entanglement only. It is necessary to investigate the robustness of the  capacities in a specific quantum information process. Our scenario to derive the MRES or MFES when an analytic proof is absent can be applied directly to these topics.


%

\begin{acknowledgments}
The authors are very grateful to the reviewers for helpful comments and
criticisms.
F.L.Z. is supported by NSF of China
(Grant No. 11105097). J.L.C. is supported by National Basic Research
Program (973 Program) of China under Grant No. 2012CB921900, NSF of
China (Grant Nos. 10975075 and 11175089) and also partly supported
by National Research Foundation and Ministry of Education, Singapore
(Grant No. WBS: R-710-000-008-271).
\end{acknowledgments}

\bibliography{MostRobFrag}

\begin{thebibliography}{33}
\expandafter\ifx\csname natexlab\endcsname\relax\def\natexlab#1{#1}\fi
\expandafter\ifx\csname bibnamefont\endcsname\relax
  \def\bibnamefont#1{#1}\fi
\expandafter\ifx\csname bibfnamefont\endcsname\relax
  \def\bibfnamefont#1{#1}\fi
\expandafter\ifx\csname citenamefont\endcsname\relax
  \def\citenamefont#1{#1}\fi
\expandafter\ifx\csname url\endcsname\relax
  \def\url#1{\texttt{#1}}\fi
\expandafter\ifx\csname urlprefix\endcsname\relax\def\urlprefix{URL }\fi
\providecommand{\bibinfo}[2]{#2}
\providecommand{\eprint}[2][]{\url{#2}}

\bibitem[{\citenamefont{Einstein et~al.}(1935)\citenamefont{Einstein, Podosky,
  and Rosen}}]{EPR}
\bibinfo{author}{\bibfnamefont{A.}~\bibnamefont{Einstein}},
  \bibinfo{author}{\bibfnamefont{B.}~\bibnamefont{Podosky}}, \bibnamefont{and}
  \bibinfo{author}{\bibfnamefont{N.}~\bibnamefont{Rosen}},
  \bibinfo{journal}{Physical Review} \textbf{\bibinfo{volume}{47}},
  \bibinfo{pages}{777} (\bibinfo{year}{1935}).

\bibitem[{\citenamefont{Nielsen and Chuang}(2000)}]{Book}
\bibinfo{author}{\bibfnamefont{M.~A.} \bibnamefont{Nielsen}} \bibnamefont{and}
  \bibinfo{author}{\bibfnamefont{I.~L.} \bibnamefont{Chuang}},
  \emph{\bibinfo{title}{Quantum Computation and Quantum Information}}
  (\bibinfo{publisher}{Cambridge University Press, Cambridge},
  \bibinfo{year}{2000}).

\bibitem[{\citenamefont{Yu and Eberly}(2004)}]{yu2004finite}
\bibinfo{author}{\bibfnamefont{T.}~\bibnamefont{Yu}} \bibnamefont{and}
  \bibinfo{author}{\bibfnamefont{J.}~\bibnamefont{Eberly}},
  \bibinfo{journal}{Physical Review Letters} \textbf{\bibinfo{volume}{93}},
  \bibinfo{pages}{140404} (\bibinfo{year}{2004}).

\bibitem[{\citenamefont{Yu and Eberly}(2006)}]{yu2006quantum}
\bibinfo{author}{\bibfnamefont{T.}~\bibnamefont{Yu}} \bibnamefont{and}
  \bibinfo{author}{\bibfnamefont{J.~H.} \bibnamefont{Eberly}},
  \bibinfo{journal}{Physical Review Letters} \textbf{\bibinfo{volume}{97}},
  \bibinfo{pages}{140403} (\bibinfo{year}{2006}).

\bibitem[{\citenamefont{Almeida et~al.}(2007)\citenamefont{Almeida, De~Melo,
  Hor-Meyll, Salles, Walborn, Ribeiro, and
  Davidovich}}]{almeida2007experimental}
\bibinfo{author}{\bibfnamefont{M.}~\bibnamefont{Almeida}},
  \bibinfo{author}{\bibfnamefont{F.}~\bibnamefont{De~Melo}},
  \bibinfo{author}{\bibfnamefont{M.}~\bibnamefont{Hor-Meyll}},
  \bibinfo{author}{\bibfnamefont{A.}~\bibnamefont{Salles}},
  \bibinfo{author}{\bibfnamefont{S.}~\bibnamefont{Walborn}},
  \bibinfo{author}{\bibfnamefont{P.}~\bibnamefont{Ribeiro}}, \bibnamefont{and}
  \bibinfo{author}{\bibfnamefont{L.}~\bibnamefont{Davidovich}},
  \bibinfo{journal}{Science} \textbf{\bibinfo{volume}{316}},
  \bibinfo{pages}{579} (\bibinfo{year}{2007}).

\bibitem[{\citenamefont{Laurat et~al.}(2007)\citenamefont{Laurat, Choi, Deng,
  Chou, and Kimble}}]{laurat2007heralded}
\bibinfo{author}{\bibfnamefont{J.}~\bibnamefont{Laurat}},
  \bibinfo{author}{\bibfnamefont{K.}~\bibnamefont{Choi}},
  \bibinfo{author}{\bibfnamefont{H.}~\bibnamefont{Deng}},
  \bibinfo{author}{\bibfnamefont{C.}~\bibnamefont{Chou}}, \bibnamefont{and}
  \bibinfo{author}{\bibfnamefont{H.}~\bibnamefont{Kimble}},
  \bibinfo{journal}{Physical Review Letters} \textbf{\bibinfo{volume}{99}},
  \bibinfo{pages}{180504} (\bibinfo{year}{2007}).

\bibitem[{\citenamefont{Konrad et~al.}(2007)\citenamefont{Konrad, De~Melo,
  Tiersch, Kasztelan, Arag{\~a}o, and Buchleitner}}]{konrad2007evolution}
\bibinfo{author}{\bibfnamefont{T.}~\bibnamefont{Konrad}},
  \bibinfo{author}{\bibfnamefont{F.}~\bibnamefont{De~Melo}},
  \bibinfo{author}{\bibfnamefont{M.}~\bibnamefont{Tiersch}},
  \bibinfo{author}{\bibfnamefont{C.}~\bibnamefont{Kasztelan}},
  \bibinfo{author}{\bibfnamefont{A.}~\bibnamefont{Arag{\~a}o}},
  \bibnamefont{and}
  \bibinfo{author}{\bibfnamefont{A.}~\bibnamefont{Buchleitner}},
  \bibinfo{journal}{Nature Physics} \textbf{\bibinfo{volume}{4}},
  \bibinfo{pages}{99} (\bibinfo{year}{2007}).

\bibitem[{\citenamefont{Yu et~al.}(2008)\citenamefont{Yu, Yi, and
  Song}}]{yu2008evolution}
\bibinfo{author}{\bibfnamefont{C.}~\bibnamefont{Yu}},
  \bibinfo{author}{\bibfnamefont{X.~X.} \bibnamefont{Yi}}, \bibnamefont{and}
  \bibinfo{author}{\bibfnamefont{H.~S.} \bibnamefont{Song}},
  \bibinfo{journal}{Physical Review A} \textbf{\bibinfo{volume}{78}},
  \bibinfo{pages}{062330} (\bibinfo{year}{2008}).

\bibitem[{\citenamefont{Li et~al.}(2009)\citenamefont{Li, Fei, Wang, and
  Liu}}]{li2009evolution}
\bibinfo{author}{\bibfnamefont{Z.~G.} \bibnamefont{Li}},
  \bibinfo{author}{\bibfnamefont{S.~M.} \bibnamefont{Fei}},
  \bibinfo{author}{\bibfnamefont{Z.~D.} \bibnamefont{Wang}}, \bibnamefont{and}
  \bibinfo{author}{\bibfnamefont{W.~M.} \bibnamefont{Liu}},
  \bibinfo{journal}{Physical Review A} \textbf{\bibinfo{volume}{79}},
  \bibinfo{pages}{024303} (\bibinfo{year}{2009}).

\bibitem[{\citenamefont{Liu and Fan}(2009)}]{liu2009dynamics}
\bibinfo{author}{\bibfnamefont{Z.}~\bibnamefont{Liu}} \bibnamefont{and}
  \bibinfo{author}{\bibfnamefont{H.}~\bibnamefont{Fan}},
  \bibinfo{journal}{Physical Review A} \textbf{\bibinfo{volume}{79}},
  \bibinfo{pages}{032306} (\bibinfo{year}{2009}).

\bibitem[{\citenamefont{Tiersch et~al.}(2008)\citenamefont{Tiersch, De~Melo,
  and Buchleitner}}]{tiersch2008entanglement}
\bibinfo{author}{\bibfnamefont{M.}~\bibnamefont{Tiersch}},
  \bibinfo{author}{\bibfnamefont{F.}~\bibnamefont{De~Melo}}, \bibnamefont{and}
  \bibinfo{author}{\bibfnamefont{A.}~\bibnamefont{Buchleitner}},
  \bibinfo{journal}{Physical Review Letters} \textbf{\bibinfo{volume}{101}},
  \bibinfo{pages}{170502} (\bibinfo{year}{2008}).

\bibitem[{\citenamefont{Vidal and Tarrach}(1999)}]{PhysRevA.59.141}
\bibinfo{author}{\bibfnamefont{G.}~\bibnamefont{Vidal}} \bibnamefont{and}
  \bibinfo{author}{\bibfnamefont{R.}~\bibnamefont{Tarrach}},
  \bibinfo{journal}{Physical Review A} \textbf{\bibinfo{volume}{59}},
  \bibinfo{pages}{141} (\bibinfo{year}{1999}).

\bibitem[{\citenamefont{Simon and Kempe}(2002)}]{PhysRevA.65.052327}
\bibinfo{author}{\bibfnamefont{C.}~\bibnamefont{Simon}} \bibnamefont{and}
  \bibinfo{author}{\bibfnamefont{J.}~\bibnamefont{Kempe}},
  \bibinfo{journal}{Physical Review A} \textbf{\bibinfo{volume}{65}},
  \bibinfo{pages}{052327} (\bibinfo{year}{2002}).

\bibitem[{\citenamefont{Zhao and Deng}(2010)}]{PhysRevA.82.014301}
\bibinfo{author}{\bibfnamefont{B.-K.} \bibnamefont{Zhao}} \bibnamefont{and}
  \bibinfo{author}{\bibfnamefont{F.-G.} \bibnamefont{Deng}},
  \bibinfo{journal}{Physical Review A} \textbf{\bibinfo{volume}{82}},
  \bibinfo{pages}{014301} (\bibinfo{year}{2010}).

\bibitem[{\citenamefont{Hill and Wootters}(1997)}]{Wootters97}
\bibinfo{author}{\bibfnamefont{S.}~\bibnamefont{Hill}} \bibnamefont{and}
  \bibinfo{author}{\bibfnamefont{W.~K.} \bibnamefont{Wootters}},
  \bibinfo{journal}{Physical Review Letters} \textbf{\bibinfo{volume}{78}},
  \bibinfo{pages}{5022} (\bibinfo{year}{1997}).

\bibitem[{\citenamefont{Wootters}(1998)}]{Wootters98}
\bibinfo{author}{\bibfnamefont{W.~K.} \bibnamefont{Wootters}},
  \bibinfo{journal}{Physical Review Letters} \textbf{\bibinfo{volume}{80}},
  \bibinfo{pages}{2245} (\bibinfo{year}{1998}).

\bibitem[{\citenamefont{\.{Z}yczkowski
  et~al.}(1998)\citenamefont{\.{Z}yczkowski, Horodecki, Sanpera, and
  Lewenstein}}]{NEG}
\bibinfo{author}{\bibfnamefont{K.}~\bibnamefont{\.{Z}yczkowski}},
  \bibinfo{author}{\bibfnamefont{P.}~\bibnamefont{Horodecki}},
  \bibinfo{author}{\bibfnamefont{A.}~\bibnamefont{Sanpera}}, \bibnamefont{and}
  \bibinfo{author}{\bibfnamefont{M.}~\bibnamefont{Lewenstein}},
  \bibinfo{journal}{Physical Review A} \textbf{\bibinfo{volume}{58}},
  \bibinfo{pages}{883} (\bibinfo{year}{1998}).

\bibitem[{\citenamefont{Lee and Kim}(2000)}]{lee2000entanglement}
\bibinfo{author}{\bibfnamefont{J.}~\bibnamefont{Lee}} \bibnamefont{and}
  \bibinfo{author}{\bibfnamefont{M.~S.} \bibnamefont{Kim}},
  \bibinfo{journal}{Physical Review Letters} \textbf{\bibinfo{volume}{84}},
  \bibinfo{pages}{4236} (\bibinfo{year}{2000}).

\bibitem[{\citenamefont{Bowen and Bose}(2001)}]{bowen2001teleportation}
\bibinfo{author}{\bibfnamefont{G.}~\bibnamefont{Bowen}} \bibnamefont{and}
  \bibinfo{author}{\bibfnamefont{S.}~\bibnamefont{Bose}},
  \bibinfo{journal}{Physical Review Letters} \textbf{\bibinfo{volume}{87}},
  \bibinfo{pages}{267901} (\bibinfo{year}{2001}).

\bibitem[{\citenamefont{Munro et~al.}(2001)\citenamefont{Munro, James, White,
  and Kwiat}}]{MEMS}
\bibinfo{author}{\bibfnamefont{W.~J.} \bibnamefont{Munro}},
  \bibinfo{author}{\bibfnamefont{D.~F.~V.} \bibnamefont{James}},
  \bibinfo{author}{\bibfnamefont{A.~G.} \bibnamefont{White}}, \bibnamefont{and}
  \bibinfo{author}{\bibfnamefont{P.~G.} \bibnamefont{Kwiat}},
  \bibinfo{journal}{Physical Review A} \textbf{\bibinfo{volume}{64}},
  \bibinfo{pages}{030302(R)} (\bibinfo{year}{2001}).

\bibitem[{\citenamefont{Wei et~al.}(2003)\citenamefont{Wei, Nemoto, Goldbart,
  Kwiat, Munro, and Verstraete}}]{MEMS1}
\bibinfo{author}{\bibfnamefont{T.-C.} \bibnamefont{Wei}},
  \bibinfo{author}{\bibfnamefont{K.}~\bibnamefont{Nemoto}},
  \bibinfo{author}{\bibfnamefont{P.~M.} \bibnamefont{Goldbart}},
  \bibinfo{author}{\bibfnamefont{P.~G.} \bibnamefont{Kwiat}},
  \bibinfo{author}{\bibfnamefont{W.~J.} \bibnamefont{Munro}}, \bibnamefont{and}
  \bibinfo{author}{\bibfnamefont{F.}~\bibnamefont{Verstraete}},
  \bibinfo{journal}{Physical Review A} \textbf{\bibinfo{volume}{67}},
  \bibinfo{pages}{022110} (\bibinfo{year}{2003}).

\bibitem[{\citenamefont{Yu and Eberly}(2002)}]{yu2002phonon}
\bibinfo{author}{\bibfnamefont{T.}~\bibnamefont{Yu}} \bibnamefont{and}
  \bibinfo{author}{\bibfnamefont{J.~H.} \bibnamefont{Eberly}},
  \bibinfo{journal}{Physical Review B} \textbf{\bibinfo{volume}{66}},
  \bibinfo{pages}{193306} (\bibinfo{year}{2002}).

\bibitem[{\citenamefont{Novotn{\`y} et~al.}(2011)\citenamefont{Novotn{\`y},
  Alber, and Jex}}]{novotny2011entanglement}
\bibinfo{author}{\bibfnamefont{J.}~\bibnamefont{Novotn{\`y}}},
  \bibinfo{author}{\bibfnamefont{G.}~\bibnamefont{Alber}}, \bibnamefont{and}
  \bibinfo{author}{\bibfnamefont{I.}~\bibnamefont{Jex}},
  \bibinfo{journal}{Physical Review Letters} \textbf{\bibinfo{volume}{107}},
  \bibinfo{pages}{090501} (\bibinfo{year}{2011}).

\bibitem[{\citenamefont{Ann and Jaeger}(2007)}]{PhysRevA.76.044101}
\bibinfo{author}{\bibfnamefont{K.}~\bibnamefont{Ann}} \bibnamefont{and}
  \bibinfo{author}{\bibfnamefont{G.}~\bibnamefont{Jaeger}},
  \bibinfo{journal}{Physical Review A} \textbf{\bibinfo{volume}{76}},
  \bibinfo{pages}{044101} (\bibinfo{year}{2007}).

\bibitem[{\citenamefont{Ali et~al.}(2008)\citenamefont{Ali, Alber, and
  Rau}}]{ali2008manipulating}
\bibinfo{author}{\bibfnamefont{M.}~\bibnamefont{Ali}},
  \bibinfo{author}{\bibfnamefont{G.}~\bibnamefont{Alber}}, \bibnamefont{and}
  \bibinfo{author}{\bibfnamefont{A.~R.~P.} \bibnamefont{Rau}},
  \bibinfo{journal}{Journal of Physics B: Atomic, Molecular and Optical
  Physics} \textbf{\bibinfo{volume}{42}}, \bibinfo{pages}{025501}
  (\bibinfo{year}{2008}).

\bibitem[{\citenamefont{Qian and Eberly}(2012)}]{Qian20122931}
\bibinfo{author}{\bibfnamefont{X.-F.} \bibnamefont{Qian}} \bibnamefont{and}
  \bibinfo{author}{\bibfnamefont{J.}~\bibnamefont{Eberly}},
  \bibinfo{journal}{Physics Letters A} \textbf{\bibinfo{volume}{376}},
  \bibinfo{pages}{2931 } (\bibinfo{year}{2012}).

\bibitem[{\citenamefont{Borras et~al.}(2009)\citenamefont{Borras, Majtey,
  Plastino, Casas, and Plastino}}]{borras2009robustness}
\bibinfo{author}{\bibfnamefont{A.}~\bibnamefont{Borras}},
  \bibinfo{author}{\bibfnamefont{A.~P.} \bibnamefont{Majtey}},
  \bibinfo{author}{\bibfnamefont{A.~R.} \bibnamefont{Plastino}},
  \bibinfo{author}{\bibfnamefont{M.}~\bibnamefont{Casas}}, \bibnamefont{and}
  \bibinfo{author}{\bibfnamefont{A.}~\bibnamefont{Plastino}},
  \bibinfo{journal}{Physical Review A} \textbf{\bibinfo{volume}{79}},
  \bibinfo{pages}{022108} (\bibinfo{year}{2009}).

\bibitem[{\citenamefont{D{\"u}r and Briegel}(2004)}]{dur2004stability}
\bibinfo{author}{\bibfnamefont{W.}~\bibnamefont{D{\"u}r}} \bibnamefont{and}
  \bibinfo{author}{\bibfnamefont{H.}~\bibnamefont{Briegel}},
  \bibinfo{journal}{Physical Review Letters} \textbf{\bibinfo{volume}{92}},
  \bibinfo{pages}{180403} (\bibinfo{year}{2004}).

\bibitem[{\citenamefont{Aolita et~al.}(2008)\citenamefont{Aolita, Chaves,
  Cavalcanti, Ac\'\i{}n, and Davidovich}}]{PhysRevLett.100.080501}
\bibinfo{author}{\bibfnamefont{L.}~\bibnamefont{Aolita}},
  \bibinfo{author}{\bibfnamefont{R.}~\bibnamefont{Chaves}},
  \bibinfo{author}{\bibfnamefont{D.}~\bibnamefont{Cavalcanti}},
  \bibinfo{author}{\bibfnamefont{A.}~\bibnamefont{Ac\'\i{}n}},
  \bibnamefont{and}
  \bibinfo{author}{\bibfnamefont{L.}~\bibnamefont{Davidovich}},
  \bibinfo{journal}{Physical Review Letters} \textbf{\bibinfo{volume}{100}},
  \bibinfo{pages}{080501} (\bibinfo{year}{2008}).

\bibitem[{\citenamefont{Zhang et~al.}(2011)\citenamefont{Zhang, Jiang, and
  Liang}}]{zhang2011speed}
\bibinfo{author}{\bibfnamefont{F.-L.} \bibnamefont{Zhang}},
  \bibinfo{author}{\bibfnamefont{Y.}~\bibnamefont{Jiang}}, \bibnamefont{and}
  \bibinfo{author}{\bibfnamefont{M.-L.} \bibnamefont{Liang}},
  \bibinfo{journal}{Arxiv preprint arXiv:1104.5057}  (\bibinfo{year}{2011}).

\bibitem[{\citenamefont{Vidal and Werner}(2002)}]{NEG1}
\bibinfo{author}{\bibfnamefont{G.}~\bibnamefont{Vidal}} \bibnamefont{and}
  \bibinfo{author}{\bibfnamefont{R.~F.} \bibnamefont{Werner}},
  \bibinfo{journal}{Physical Review A} \textbf{\bibinfo{volume}{65}},
  \bibinfo{pages}{032314} (\bibinfo{year}{2002}).

\bibitem[{\citenamefont{\.{Z}yczkowski and Ku\'{s}}(1994)}]{randomU}
\bibinfo{author}{\bibfnamefont{K.}~\bibnamefont{\.{Z}yczkowski}}
  \bibnamefont{and} \bibinfo{author}{\bibfnamefont{M.}~\bibnamefont{Ku\'{s}}},
  \bibinfo{journal}{Journal of Physics A: Mathematical and General}
  \textbf{\bibinfo{volume}{27}}, \bibinfo{pages}{4235} (\bibinfo{year}{1994}).

\bibitem[{\citenamefont{Verstraete et~al.}(2001)\citenamefont{Verstraete,
  Audenaert, Dehaene, and Moor}}]{verstraete2001comparison}
\bibinfo{author}{\bibfnamefont{F.}~\bibnamefont{Verstraete}},
  \bibinfo{author}{\bibfnamefont{K.}~\bibnamefont{Audenaert}},
  \bibinfo{author}{\bibfnamefont{J.}~\bibnamefont{Dehaene}}, \bibnamefont{and}
  \bibinfo{author}{\bibfnamefont{B.}~\bibnamefont{Moor}},
  \bibinfo{journal}{Journal of Physics A: Mathematical and General}
  \textbf{\bibinfo{volume}{34}}, \bibinfo{pages}{10327} (\bibinfo{year}{2001}).

\end{thebibliography}

\end{document}